\documentclass[12pt,  amssymb]{iopart}
\usepackage{iopams}
\usepackage{graphicx} 
\usepackage{bm} 
\usepackage[usenames,dvipsnames]{color}

\newcommand{\ket}[1]{| #1 \rangle}

\newcommand{\beq}{\begin{equation}}
\newcommand{\eeq}{\end{equation}}

\def\xi{\mathbf{x}_i}

\def\vp{\Phi}
\def\D{\Delta}

\newcommand {\tz}[1]{ {\textcolor{OliveGreen}{#1}}}           

\begin{document}

\title{Non-adiabatic control of quantum energy transfer in ordered {and disordered} arrays}

\author{Ping Xiang, Marina Litinskaya, Evgeny A. Shapiro, and Roman V. Krems}

\address{Department of Chemistry, University of British Columbia, Vancouver, V6T 1Z1, Canada}

\begin{abstract}
An elementary excitation in an aggregate of coupled particles
generates a collective excited state. We show that the dynamics of
these excitations can be controlled by applying a transient external
potential which modifies the phase of the quantum states of the 
individual particles. The method is based on an interplay of
adiabatic and sudden time scales in the quantum evolution of the many-body states. We show that
specific phase transformations can be used to accelerate or
decelerate quantum energy transfer and spatially focus delocalized
excitations onto different parts of  arrays of quantum
particles. {We consider possible experimental implementations of
the proposed technique and study the effect of disorder due to
the presence of impurities on its fidelity. {We further show
that the proposed technique can allow control of energy transfer  in completely disordered systems.}}
\end{abstract}

\maketitle

\section{Introduction}

The experiments with ultracold atoms and molecules trapped in optical lattices have opened a new frontier of condensed-matter physics research. The unique properties of these systems -- in particular, large ($>$ 400 nm) separation of lattice sites, the possibility of tuning the tunnelling amplitude of particles  between lattice sites by varying the trapping field and the possibility of controlling interparticle interactions with external electric or magnetic
 fields -- offer many exciting applications ranging from quantum simulation of complex lattice models \cite{Barnett2006, micheli2006, Brennen2007, Buchler2007, Carr1, Carr2, Trefzger2010, Kestner2011, gorshkov1, gorshkov2}
 to the study of novel quasi-particles \cite{biexcitons} that cannot be realized in solid-state crystals. In the limit of strong trapping field, ultracold atoms or molecules
 on an optical lattice form a Mott insulator phase, in which each lattice site is populated by a fixed number of particles. With one particle per lattice site, this phase represents a periodic,
crystal-like structure. Such a system can be thought of as a prototype  of an ordered system, in which a single lattice site (or a small number of lattice sites) can be individually addressed
by an external field of a focused laser beam. This can be exploited for engineering the properties of quantum many-body systems by changing the energy of particles in individual lattice sites \cite{our-2012-prl}.

In the present work, we consider the generic problem of energy transfer -- i.e. the time evolution of an elementary  
quantum excitation -- in such a system. In particular, we explore the possibility of controlling energy transfer through
 an array of coupled quantum monomers by applying monomer-specific external perturbations. {This is necessary for 
several applications.} First, collective excitations in molecular arrays in optical lattices have been proposed as 
high-fidelity candidates for quantum memory \cite{rabl}.  The ability to manipulate collective excitations is necessary
 for building scalable quantum computing networks \cite{quantum-computing}. Second, ultracold atoms and 
molecules in optical lattices can be perturbed by a disorder potential with tunable strength 
\cite{anderson-localization-of-ultracold-atoms}. Engineering localized and delocalized excitations in such systems 
can be used to investigate the role of disorder-induced perturbations  on quantum energy transfer, a question of
 central importance for building efficient light-harvesting devices \cite{solar-cell}.
 {Third, the possibility of controlling energy transfer in an optical lattice with ultracold atoms or molecules can
 be used to realize inelastic scattering processes with both spatial and temporal control. Finally, control over energy
transfer in quantum systems can be used for studying condensed-matter excitations and energy transport without 
statistical averaging.}

An excitation of a coupled many-body system generates a wave packet representing a coherent superposition of single-particle excitations. The method proposed here is based on shaping such many-body wave packets by a series of sudden perturbations, in analogy with the techniques developed for strong-field alignment and orientation of molecules in the gas phase \cite{alignment-review}. Alignment is used in molecular imaging experiments and molecular optics \cite{alignment-review, imaging1, imaging2, imaging3}, and is predicted to provide control over mechanical properties of molecular scattering \cite{averbukh-AlignedInteractions, averbukh-AlignedInteractions2}. Here, we consider the use of similar techniques for controlling quantum energy transfer in a many-body system.  {When applied to a completely ordered system,} the proposed method is reminiscent of the techniques used to move atoms in optical lattices, where a uniform force is applied for a short period of time \cite{denschlag2002}. The conceptual difference comes from the fact that in the present case the momentum is acquired by a quasi-particle -- a collective excitation distributed over many monomers. During the subsequent evolution, the particles do not move -- rather, the excitation is transferred from one monomer to another. In order to control such excitations, we exploit an interplay of the adiabatic and sudden time scales, which correspond to single-monomer and multi-monomer evolution. We also exploit the wave-like nature of the excitation wave function to draw on the analogy with wave optics. This analogy, too, is not complete due to the discrete nature of the lattice.

In order to emphasize the generality of the proposed method, we formulate the problem and present the results
 in Sections 2  and 3 in terms of the general Hamiltonian parameters. Section 4 then describes how the external 
perturbations corresponding to the results presented can be realized in experiments with ultracold atoms and
 molecules. Section 5 discusses controlled energy transfer in systems with, specifically, 
dipole - dipole interactions. Section 6 considers the effects of lattice vacancies on the possibility of controlling 
energy transfer and {Section 7 extends the proposed technique to
control of excitation dynamics in strongly disordered arrays with a large
concentration of impurities. Section 8 presents the conclusions.}

\section{Sudden phase transformation}
Consider, first, an ensemble of ${N}$ coupled identical monomers
possessing two internal states arranged in a one-dimensional array
with translational symmetry. The Hamiltonian for such a system is
given by
\begin{equation}
H_{\rm{exc}} = \Delta E_{e-g}\sum_n | e_n \rangle \langle e_n | +
\sum_{n,m} \alpha(n-m) |e_n, g_m\rangle \langle g_n, e_m | \ ,
\label{ham}
\end{equation}
where  $|g_n \rangle$ and $|e_n\rangle$ denote the ground and
excited states in site $n$, $\Delta E_{e-g}$ is the monomer
excitation energy and $\alpha(n-m)$ represents the coupling
between two monomers at sites $n$ and $m$.
The singly excited state of the system is
\begin{eqnarray}
|\psi_{\rm exc}\rangle = \sum_{n=1}^N C_n |e_n\rangle \prod_{i
\neq n} |g_i\rangle . \label{basis}
\end{eqnarray}
In general, the expansion coefficients $C_n$ are complicated functions of $n$ determined by the properties of the system, in particular, the translational invariance or lack thereof as well as the strength of disorder potential. 
 If an ideal, periodic system with lattice constant $a$ is excited by a single-photon transition, 
the expansion coefficients are  $C_n =
e^{iakn}/\sqrt{N}$ and  $|\psi_{\rm exc} \rangle \Rightarrow
|\psi_{\rm exc}(k)\rangle$ represents a quasi-particle called
Frenkel exciton, characterized by the wave vector $k$
\cite{agranovich}. The magnitude of the wave vector $k$ is determined by the conservation of the total (exciton plus photon) momentum. 
The energy of the exciton is given by $E(k) =
\Delta E_{e-g} + \alpha(k)$ with $\alpha(k) =\sum_n \alpha(n)
e^{-i  a k n}$. In the nearest neighbor approximation,
\begin{eqnarray}
E(k) = \Delta E_{e-g} + 2 \alpha \cos ak, \label{dispersion}
\label{Eexc}
\end{eqnarray}
where $\alpha=\alpha(1)$.

%

With atoms or molecules on an optical lattice, it is also possible to generate a localized excitation placed on a single site
 (or a small number of sites) by applying a gradient of an external electric or magnetic field and inducing transitions in
 selected atoms by a pulse of resonant electromagnetic field \cite{demille}. The presence of a disorder potential, 
whether coming from jitter in external fields or from incomplete population of lattice sites,  also results in spatial 
localization. Similar to how Eq.~(\ref{basis}) defines the collective excited states in the basis of lattice sites, 
any localized excitation $| \psi \rangle$ can be generally written as a coherent superposition of the exciton states
 $| \psi_{\rm exc}(k)\rangle$ with different $k$:

\begin{eqnarray}
| \psi \rangle = \sum_{k} G_k | \psi_{\rm exc}(k) \rangle.
\label{k-rep}
\end{eqnarray}

Control over energy transfer in an ordered array can be achieved
by (i) shifting the exciton wave packets in the momentum
representation (which modifies the group velocity and the shape
evolution of the wave packets) and (ii) focusing the wave packets
in the coordinate representation to produce localized excitations
in an arbitrary part of the lattice. To achieve this, we propose
to apply a series of site-dependent perturbations  that modify the phases of the quantum states of spatially separated monomers. 
These phase transformations change the dynamics of the time evolution of the collective excitations. 
Here we consider the transformations leading to acceleration or deceleration of collective excitations, while   
 the focusing phase transformations are described in Section 3.

For modifying the group velocity of a collective excitation, the
essential idea is to add a factor $e^{i \delta a n}$ to each term
in the expansion (\ref{basis}), so that each $| \psi_{\rm exc}(k)
\rangle$ component in a wave packet is transformed into $|
\psi_{\rm exc}(k+\delta)\rangle$. This transformation shifts the
wave packets by $\delta$ in $k$-space while preserving their
shape. As a result, one can engineer wave packets probing any part
of the dispersion $E(k)$ leading to different group velocity and
shape evolution. The feasibility of such transformation in an ensemble of atoms or molecules on an optical lattice is discussed below and in Section 4. 

Adding a site-dependent phase to the excitonic wavefunction
exploits an interplay of the adiabatic and sudden time scales.
Consider  the $n$-th monomer subjected to an external field
$\mathcal{E}_n(t)$ which varies from 0 to some value and then back
to 0 in time $T$. If the variation is adiabatic with respect to
the evolution of the free monomer states, $T\gg \hbar/\Delta
E_{e-g}$, each eigenstate $|f\rangle$ of the monomer acquires a
state-dependent phase shift \cite{adiabatic-theo}
\begin{eqnarray}
\ket{f_n (T)} = e^{-i \phi_{n}^f}\ket{f_n(0)},
\label{adiabatic-theorem} \label{phase}
\end{eqnarray}
where $\phi_{n}^f = \frac{1}{\hbar}\int_{0}^{T} E_{n}^f(t )dt$,
$E_{n}^f (t)$ is the instantaneous eigenenergy and $f$ can be $e$
or $g$.
Now consider the action of such phase change on the collective
excitation state (\ref{basis}). If $T\ll \hbar/{\alpha}$, the
change is sudden with respect to the excitation transfer between
monomers and the state (\ref{basis}) acquires a site-dependent
phase $\Phi_n = \phi_{n}^e-\phi_{n}^g$. If $\Phi_n = \Phi_0 + n a
\delta $, then the momentum $\delta$ is imparted onto the
excitonic wavefunction. By analogy with ``pulsed alignment of
molecules'' \cite{alignment-review}, we call this transformation a
``phase kick'' or ``momentum kick''. Its action is also similar to
that of a thin prism on a wavefront of a monochromatic laser beam.

\begin{figure}[ht]
\centering
\includegraphics[width=\linewidth]{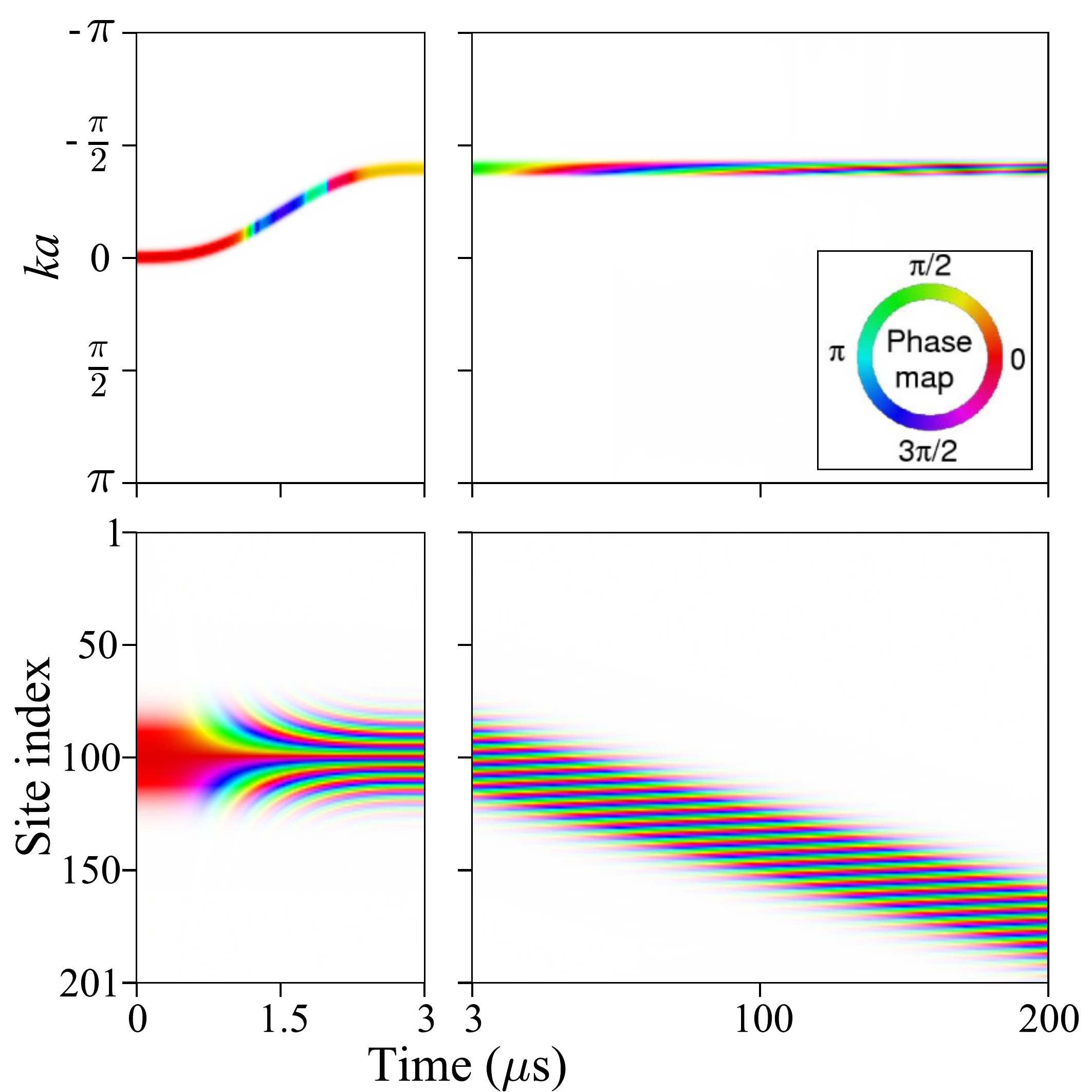}
\caption{(Color online) Evolution of the exciton wave packet in
the momentum and coordinate spaces. The phase of the wave function
is shown by color. The calculation is for a one-dimensional array
of 201 monomers with $\alpha = 22.83$ kHz and $\Delta E_{e-g} =
12.14$ GHz, which corresponds to LiCs molecules trapped on an
optical lattice with lattice constant $a = 400$ nm and subjected
to a homogeneous DC field of 1 kV/cm directed perpendicular to the
intermolecular axis (for details, see Section 4). The kicking potential leading to
 a phase transformation $\Phi_n \simeq \Phi_0  -1.29 n $ is provided by a
$\lambda = 1064$ nm Gaussian laser beam, with the propagation
direction along the array axis, focused to 5 $\mu$m, with the
intensity at the focus equal to $10^7$ W/cm$^2$. The laser pulse
is on between 0 and  $3$ $\mu$s. The molecules are placed on the
beam axis with the first molecule 5 $\mu$m away from the focus.
} \label{momentum-kick}
\end{figure}

In order to illustrate the shifting of exciton wave packets in the
momentum space, we solve numerically the time-dependent
Schr\"{o}dinger equation with the unperturbed Hamiltonian
(\ref{ham}),  subjected to a transient site-dependent external
perturbation that temporarily modulates $\Delta E_{e-g}$.
We choose the parameters $\Delta E_{e-g}$, $\alpha$ and the lattice constant $a$ that correspond to an array of
polar molecules trapped in an optical lattice, as described in Section 4.
The time-dependent perturbation, chosen to vary almost linearly along the lattice, has the form of a short pulse with the duration $T = 3\; \mu s$. 
The phase acquired by the particles during this time is given by $\Phi_n \simeq \Phi_0  -1.29 n $, which can be achieved with a focused laser beam, as described in Section 4.


The excitation at $t=0$ is described by
a Gaussian wave packet of the exciton states $|\psi_{\rm
exc}(k)\rangle$, with the central wavevector $k=0$.
Fig.~\ref{momentum-kick} shows that the
entire wave packet acquires momentum during the external perturbation pulse (left
panels). This is manifested as a phase variation in the coordinate
representation, and as a shift of the central momentum in the
$k$-representation. After the external perturbation is gone, the
wave packet does not evolve in the $k$-representation and moves
with the acquired uniform velocity in the coordinate
representation.



\section{Focusing of a delocalized excitation}

In order to achieve full control over excitation transfer, it is desirable to find a particular phase transformation
that focuses a delocalized many-body excitation onto a small part of the lattice, ideally a single lattice site.
In optics, a thin lens focuses a collimated light beam by shifting
the phase of the wavefront, thus converting a plane wave to a
converging spherical wave. Similarly, a phase kick can serve as a
time domain ``lens'' for collective excitations: an excitation initially
delocalized over a large number of monomers can be focused onto a
small region of the array after some time. By analogy with optics, a concave
symmetric site-dependent phase $\Phi(n)$ applied simultaneously to
all monomers may turn a broad initial distribution $C_n(t=0)$ into
a narrow one.

The dynamics of the excitation state in the lattice is determined by the time dependence
of the coefficients $C_n(t)$ in Eq. (\ref{basis}). In order to find the expression for $C_n(t)$, we expand the amplitudes
 at $t=0$ in a Fourier series
\begin{eqnarray}
C_n(t=0) = \sum_q
\frac{e^{iqn}}{\sqrt{N}}C(q; t=0)
\end{eqnarray}
and apply the propagator $e^{-iE(q)t/\hbar}$ to each $q$-component with
$E(q)$ representing the exciton energy given by Eq.~(\ref{dispersion}).
Transforming the amplitudes $C(q)$ back to the site representation then yields
\begin{equation}
C_m(t) = \frac{1}{N} \sum_{n,k} C_n(t=0) e^{i [\Phi(n) +  k a
(m-n) - E(k) t/\hbar ]},
 \label{Cn(t)}
\end{equation}
where $\Phi(n)$ is a
site-dependent phase applied at $t=0$, as described in the previous section.
Note that the phase $\Phi(n)$ does not have to be applied
instantaneously. The phase $\phi(n, \tau)$ can be applied continuously over an
extended time interval as long as the accumulated phase gives the desired outcome $\int_0^{T}
\phi(n, \tau)d\tau = \Phi(n)$.

As Eq. (\ref{Cn(t)}) shows, the focusing efficiency is determined by the phase transformation and
the shape of the dispersion curve $E(k)$. Given the cosine dispersion of excitons (\ref{dispersion}), is it possible to
 focus a delocalized excitation onto a single lattice site?
To answer this question, we assume that $C_{n = n_0}(t=0) = 1$ and apply Eq. (\ref{Cn(t)}) to
 calculate the coefficients $C_m(t)$ at $t=-\tau$.  Using
the expansion of an exponent in Bessel functions
\begin{eqnarray}
e^{-ia\cos x} = \sum_{n} e^{i(x-\pi/2)n} J_n(a)
\end{eqnarray}
and the orthonormality  of the Bessel functions
\begin{eqnarray}
\sum_{n} J_n(x) J_{n-m}(x) = \delta_{m,0},
\end{eqnarray}
we find that the wave packet (\ref{basis}) with the expansion coefficients
\begin{equation}\label{C_ideal}
C_n^{(n_0)} = J_{n-n_0}(2\alpha \tau) e^{i \pi (n-n_0) /2}
\end{equation}
focuses, upon coherent evolution, in time $\tau$ on a single site $n_0$.
This shows that a phase transformation alone is, generally,  not sufficient to create a collective excitation state that focuses
onto a single lattice site.
The best focusing must involve both the phase and amplitude modulations, which may be difficult to realize in
experiments. A simpler procedure can be implemented if the phase transformations are restricted to a particular part of
the exciton dispersion.

From wave optics, waves with quadratic dispersion can be
focused, while those with linear dispersion propagate without changing
the wave packet shape \cite{focusing-books-1,focusing-books-2}. It is this interplay of the quadratic (at low $k$) and linear (at $k \approx \pm \pi/2a$)
parts of the cosine-like exciton dispersion (\ref{dispersion}) that precludes perfect focusing of a general collective excitation.
In order to avoid the undesirable amplitude modulations, it may be possible to focus delocalized excitations by a phase transformation that
 constrains the wave packet (\ref{k-rep}) to the quadratic part of the dispersion $E(k)$.  For such wave packets, adding a quadratic phase
$\vp(n) = \vp_0 (n-n_0)^2$ must lead to
focusing around site $n_0$. Below we illustrate the effect of the quadratic phase transformation for two types of initial states.

First, consider a broad Gaussian wave packet (\ref{basis}) with
$C_n(\tilde\sigma_x;t=0) =\sqrt{{a}/{\tilde\sigma_x \sqrt{\pi}} }$
$ \exp \left[- {a^2 (n-n_0)^{2}} / {2\tilde\sigma_{x}^2 } \right]$
where $\tilde\sigma_x \gg a$ {is the initial width. The corresponding width
in the wave vector space is given by $\sigma_k = 1/\tilde\sigma_x$.
The application of an inhomogeneous phase $\Phi(n) = \vp_0
(n-n_0)^2$ at $t=0$ results in additional broadening of the
initial state, and the total width of the wave packet in the
wave vector space with the account of the phase-induced contribution
becomes}
\cite{focusing-books-1,focusing-books-2} 
%
\begin{equation}
\sigma_{k}(\tilde\sigma_x, \Phi_0)  =
\frac{1}{\tilde\sigma_x}\,{\sqrt{1 + 4 \Phi_0^2 \tilde\sigma_x^4 /
a^4 }}. \label{sigma-k}
\end{equation}
%

{By analogy with optics, one should expect better focusing
with larger $\Phi_0$ (the width of the wave packet in real space
is $\sigma_x(\Phi_0) = 1/\sigma_k(\tilde\sigma_x,\Phi_0))$.}
However, large values of $\Phi_0$ may take the wave packet outside the quadratic
part of the dispersion, impeding the focusing.
To find the optimal phase $\Phi_0^*$ that keeps the wave packet
within the quadratic dispersion while focusing it, we use the
condition $\Delta_k = a \sigma_k \lesssim 1$, which yields
$\Phi_0^* = \pm a/2 \tilde\sigma_x$ for the optimal focusing. At
time
\begin{equation}
t_* \approx 1 / 4 \alpha \Phi_0^* \ , \label{focus-time}
\end{equation}
the wave packet is most
focused and has a width
\begin{equation}
\sigma_{x,F} (\Phi_0^*) =  \frac{\tilde\sigma_x} {\sqrt{1 + 4
\Phi_0^{*\,2} \tilde\sigma_x^4 / a^4 }} \approx a .
\label{x-focusing-Gauss}
\end{equation}
For the time $t_*$ in Eq. (\ref{focus-time}) to be positive, $\alpha$ and $\Phi_0^*$ must have the same sign. Therefore, a convex quadratic phase profile
$\Phi(n)$ with $\Phi_0>0$ must focus collective excitations in a system with
repulsive couplings between particles in different lattice sites ($\alpha>0$), and a concave quadratic phase
profile $\Phi(n)$ with $\Phi_0<0$ must focus excitations in a system with
attractive couplings ($\alpha<0$).

Second, consider a completely delocalized excitation (\ref{basis})
with $C_n(k;t=0) = {e^{iakn}}/{\sqrt{N}}$ describing an eigenstate
of an ideal system of $N$ coupled monomers.
If $E(k)$ in Eq.~(\ref{Cn(t)}) is approximated as $E(k) = \D E_{e-g}
-\alpha a^2 k^2$, the quadratic phase transformation $\vp(n) = \vp_0
n^2$ yields

\begin{equation}
\begin{array}{c}
\displaystyle C_m(t) = \frac{e^{-i\alpha a^2 k^2}}{N}\
\sqrt{\frac{i\pi}{N \Phi_0}} \sum\limits_q e^{i[a^2(k-q)^2 (\alpha
t - 1/4\Phi_0) + q
a (m + 2\alpha ak) ]}\times\\

\\

\displaystyle\times\Theta\left(-\frac{N\Phi_0}{a} < k-q <
\frac{N\Phi_0}{a}
\right),\\
\end{array}\label{C_plane_wave}
\end{equation}
where $\Theta(z) = 1$ if $z$ is true and zero otherwise.
In order to derive Eq. (\ref{C_plane_wave}), we
used the approximate equality
\begin{equation}\label{approx integral}
\int\limits_{-M}^M dx\ e^{-i(ax^2 + b x)} \approx
\sqrt{\frac{\pi}{ia}}\ e^{i b^2/4a}\  \Theta(-2Ma < b < 2Ma),
\end{equation}
obtained by approximating the error function of a complex argument
${\rm Erf}(\sqrt{i}x)$ by the sign function, which is accurate for
large argument $x$.

At time $t_* = 1/4\alpha\Phi_0$, the terms quadratic in $q$
in Eq.~(\ref{C_plane_wave}) are canceled, and the sum over $q$ reduces to a
delta-function, if the summation limits are from $-\pi/a$ to
$\pi/a$. Therefore, the choice $\Phi_0 = \pi / N$ yields $C_m(t) =
\sqrt{i} e^{-i N a^2 k^2 / 4\pi} \delta_{m,-\nu_k}$, where   $\nu_k$
is the index of the initial wave vector $k = 2\pi \nu_k / Na$, quantized due to the discreteness of the lattice.
According to Eq.~(\ref{C_plane_wave}), the dimensionless width
of the wave packet in the wave vector space is $\Delta_k(\Phi_0)
\equiv a\sigma_{k}(\Phi_0) \approx 2N\Phi_0$. When $\Phi_0 = \pi /
N$, the wave packet spreads over the entire Brillouin zone, including the
linear parts of the exciton dispersion.
Using Eq.~(\ref{approx integral}) we find that for an arbitrary
value of $\Delta_{k}(\Phi_0)$, the site amplitudes at the time of
focusing $t_* = 1/4\alpha\Phi_0$ are
\begin{equation}
C_n(k; t=t_*) \approx \frac{e^{i\Delta(\Phi_0) n^2/2N}}{n}
\sqrt{\frac{2i}{\pi \Delta_{k}(\Phi_0)}} \sin (n
\Delta_{k}(\Phi_0) /2). \label{x-focusing-PlaneWave}
\end{equation}
In order to keep the linear part of the dispersion spectrum
unpopulated, we choose the optimal focusing phase $\Phi_0^*
\sim 1 /2N$, so that $\Delta_{k}( \Phi_0^*)\sim 1$.

\begin{figure}[ht]
\centering
\includegraphics[width=\linewidth]{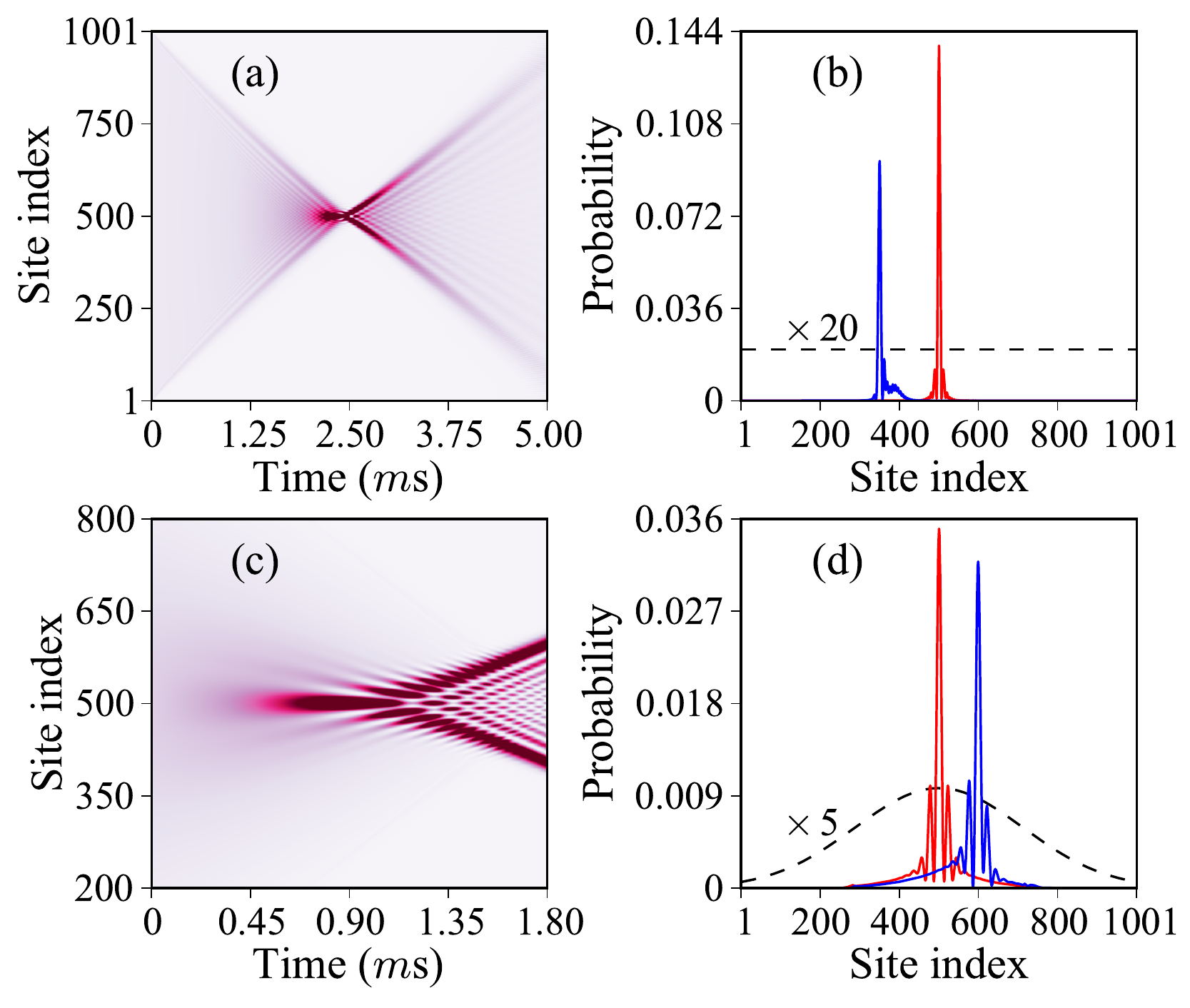}
\caption{ Focusing of a completely delocalized collective
excitation (panels a and b) and a broad Gaussian wave packet of
Frenkel excitons (panels c and d) using the quadratic phase
transformations at $t=0$ as described in text. The dashed lines
show the initial distribution magnified by 20 and 5 respectively
in (b) and (d). The solid curves in panels (b) and (d) correspond
to two different phase transformation focusing the same wave
packet onto different parts of the array.  The calculations are
performed with the same parameters $\alpha$, $a$, and $\Delta
E_{e-g}$ as in Figure \ref{momentum-kick}. The results are computed
with all couplings accounted for. }\label{focusing-1d}
\end{figure}

Eqs. (\ref{x-focusing-Gauss}) and (\ref{x-focusing-PlaneWave}) are
valid for a many-body system with nearest neighbor interactions
only. In most physical systems, the energy dispersion is modified
by long-range couplings. In order to confirm that the above
predictions are also valid for systems with long-range
 interactions and illustrate the focusing of
delocalized excitations, we compute the time evolution of the wave
packets by solving the wave equation numerically for a system with
long-range dipole-dipole interactions. Figure \ref{focusing-1d}
illustrates the focusing dynamics of a completely delocalized
excitation (panels a and b) and a broad Gaussian wave packet (panels c and d) in a
system with all (first neighbour, second neighbour, etc.) couplings explicitly included  in the
calculation. The results show that  
the collective excitations can be focused to a few lattice sites. The role of the long-range coupling will be
explicitly discussed in Section 6.

The focusing scheme demonstrated above can be generalized to  systems of higher
dimensionality. To illustrate this, we repeated the calculations
presented in Figures \ref{focusing-1d}c and \ref{focusing-1d}d for
a delocalized excitation placed in a square 2D lattice with an external
potential that modulates the phase as a function of both $x$ and
$y$. Figure 3 shows the focusing of an initially broad wave packet
onto different parts of a 2D lattice induced by the quadratic
phase transformation $\Phi(x,y) = \Phi_0[ (n_x - n_{x_0})^2 + (n_y- n_{y_0})^2]$, 
where $n_x$ and $n_y$ are the lattice site indices along the $x$ and $y$ directions. 
The calculations include all long-range couplings as in Figure \ref{focusing-1d}. The comparison of Figures 2(c,d) and 3 illustrates that the 
focussing efficiency in 2D is greater. The results also demonstrate that the delocalized excitations can be effectively focused on different parts of the lattice simply by varying the reference site $(n_{x_0}, n_{y_0})$ in the phase transformation. 

\begin{figure}[ht]
\centering
\includegraphics[width=\linewidth]{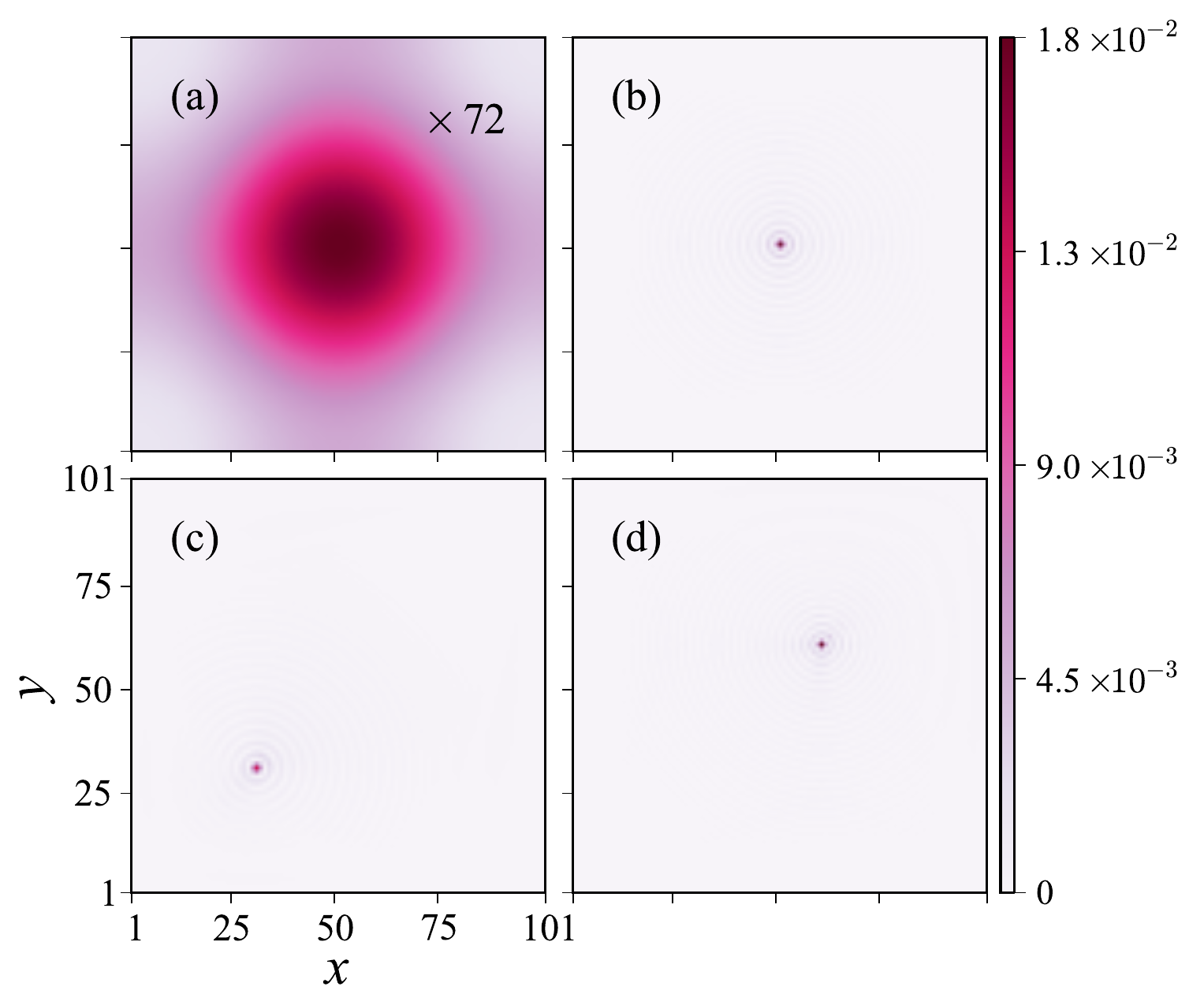}
\caption{Focusing of a delocalized excitation in a 2D array shown
at $t=0$ in panel (a) onto different parts of the lattice (panels
b--d). The probability distribution in panel (a) is enhanced by
the factor 72 for clarity.  The calculations are performed with
the same parameters $\alpha$, $a$, and $\Delta E_{e-g}$ as in
Figure \ref{momentum-kick} and the quadratic phase transformation
at $t=0$.
 }\label{focusing-2d}
\end{figure}

\section{Controlled excitations of ultracold atoms and molecules}

The techniques proposed in Sections 2 and 3 can be realized with ultracold atoms or molecules trapped in an
optical lattice in a Mott insulator phase \cite{Ye-arrays-PRL12}.
There are three general requirements that must be satisfied:

\begin{itemize}
\item
 (i) The time required for a simple phase transformation must
be shorter than the spontaneous decay time of the excited
state.

\item
 (ii) The overall coherence of the system  must be preserved
on the time scale of the excitonic evolution in the entire array, set by $K\hbar/\alpha$, where $K$
 is the number of monomers participating in the dynamics of the collective excitation.

\item
(iii) The lattice constant must be large enough to allow considerable variation of the external
 perturbation from site to site.

\end{itemize}

Optical lattices offer long coherence times ($> 1$ sec) and large lattice constants ($>400$ nm)
 \cite{optical-lattice-review}. The lifetime of the
collective excitations depends on the internal states
 of the particles used in the experiment and the momentum distribution of the excitonic states in the wave packet
(\ref{k-rep}).

For ultracold alkali metal atoms in an optical lattice, an optical excitation may generate collective states (\ref{basis}),
as discussed in Refs. \cite{Zoubi1, Zoubi2, Zoubi3, Zoubi4, Zoubi5, Zoubi6, Zoubi7, Zoubi8, Zoubi9}.
The lifetime of these excited states is limited by the spontaneous emission of the electronically excited atoms and is
in the range of 10 - 30 ns.
However, the collective excited states can be protected from spontaneous emission if the wave vector range populated by excitons in the wave packet (\ref{k-rep}) is
outside of the light cone, so that $k >\Delta E_{e-g}/\hbar c$ \cite{agranovich}. These states do not readily radiate as energy and momentum conservation cannot be simultaneously satisfied for them. The emission of photons may occur on a much longer time-scale at the array boundaries or due to perturbations breaking the translational symmetry. Due to the same conservation laws,
single-photon excitation of atomic ensembles always generates excitons with $k \approx 0$.
Once these excited states are created, the phase-kicking technique introduced in Section 2 can be used to shift the excited states in the wave vector space
away from $k=0$ (cf. Figure 1) and thus protect the excited states from fast spontaneous decay. This phase transformation can be induced by a pulse of an off-resonant laser field $\mathcal{E}_{AC}$,
detuned from the  $e \leftrightarrow g$ resonance by the value  $\delta\omega$, leading to the
AC Stark shift {(see e.g. Ref.\cite{focusing-books-1})}
\begin{equation}
\Delta E_{AC} =  \mathcal{E}_{AC}^2 \frac{V_{eg}^2}{4 \delta \omega} \ ,
\end{equation}
where $V_{eg}$ is the matrix element of the dipole-induced transition.   By choosing $ V_{eg}= 1$ a.u.,
$\delta \omega = 3 V_{eg}$, and the laser intensity $I = 5\times10^{10}$ W/cm$^2$, we obtain that the shift
 $\phi = \D E_{AC} \times T_{\rm pulse} = \pi $ can be achieved in less than 1 ns.

{This shift brings a wave packet initially centered at $k=0$ to the ``dark" edge of the Brillouin zone, where the dispersion of excitons is still quadratic and all the focusing schemes discussed in Section 3 can be applied. 
Another phase transformation can bring the excited state back to the $k\approx 0$ region, where it can be observed via fast spontaneous emission. 
 The experiments with ultracold atoms have demonstrated the creation of a Mott insulator phase with the lattice filling factor reaching 99 \%  \cite{atom-mott1, atom-mott2, atom-mott3}.
 The phase transformations proposed here can be used to stabilize excitonic states in ultracold atomic ensembles against spontaneous emission  
for multiple interesting applications \cite{Zoubi1, Zoubi2, Zoubi4,  Zoubi6,  Zoubi8}. 


The spontaneous decay problem can be completely avoided by using rotational excitations in an ensemble of
ultracold polar molecules trapped in an optical lattice.
The rotational states are labeled by the quantum
number of the rotational angular momentum $\bm{J}$ and the
projection $M_J$ of $\bm{J}$ on the space-fixed quantization axis
$Z$. We choose the rotational ground state $|J=0, M_J=0\rangle$ as
$|g\rangle$ and the rotational excited state $|J=1, M_J = 0
\rangle$ as $|e\rangle$. The state $|J=1, M_J = 0 \rangle$ is
degenerate with the states $|J=1, M_J = \pm 1 \rangle$. This
degeneracy can be lifted by applying a homogeneous DC electric
field, making the $|g\rangle$ and $|e\rangle$ states an isolated
two-level system. The molecules in different lattice sites are
coupled by the dipole-dipole interaction $V_{\rm dd}(n-m)$.
The magnitude of
the coupling constant $\alpha(n-m) = \langle e_{n} , g_{m} |
{V}_{\rm dd}(n-m) | g_{n}, e_{m} \rangle$ between molecules with
the dipole moment $1$ Debye separated by 500 nm is on the order of
1 kHz \cite{felipe}. Due to the low value of $\Delta E_{e-g}$, the spontaneous emission time of
rotationally excited molecules  exceeds 1 second.


For molecules on an optical lattice, one can implement the phase
kicks by modifying the molecular energy levels with pulsed AC or
DC electric fields. The rotational energy levels for $^1\Sigma$
molecules in a combination of weak AC and DC electric fields are
given by \cite{friedrich-95}
\begin{eqnarray}
E_{J,M_J}  &\approx&  BJ(J+1) +  \frac{\mu^2 \mathcal{E}_{DC}^2
}{2B} G(J,M_J) \nonumber \\
&& - \frac{ \alpha_{\perp}\mathcal{E}_{AC}^2 }{4} + \frac{
(\alpha_{||}-\alpha_{\perp})\mathcal{E}_{AC}^2 }{4}F(J,M_J)
 \label{DressedEnery-Sum}
\end{eqnarray}
where $B$ is the rotational constant,  $G(0,0) = -1/3$, $G(1,0) =
1/5$, $F(0,0) = -1/3$, $F(1,0) = -3/5$, $\mathcal{E}_{AC}$ is the
envelope of the quickly oscillating AC field, $\alpha_\|$ and
$\alpha_{\perp}$ are the parallel and perpendicular
polarizabilities and $\mu$ is the permanent dipole moment of the
molecule.

The momentum shift of the exciton wave packets can be achieved by
applying a time-varying DC electric field $\mathcal{E}(t)
=\mathcal{E}_{\ast} + \mathcal{E}(n)\sin^{2}(\pi t/T)$, where
$\mathcal{E}(n)$ is linear with respect to $n$. Assuming that
$\mathcal{E}(n) = (n-n_0) A$ and $\mathcal{E}(n) \ll
\mathcal{E}_{{\ast}}$, and using Eqs. (\ref{phase}) and
(\ref{DressedEnery-Sum}), gives  $\delta = 4 A
\mathcal{E}_{{\ast}} \mu^2 T / 15 \hbar B a$. We have confirmed
this result by a numerical computation showing that for LiCs
molecules in an electric field of $\mathcal{E}_{{\ast}}=1$ kV/cm,
an electric field pulse with $A=7.434\times 10^{-4}$ kV/cm and
$T=1$ $\mu$s results in a kick of $\delta = \pi/2 a$, bringing an
excitonic wave packet from the $k=0$ region to the middle of the
dispersion zone.

An alternative strategy is to use a pulse of an
off-resonance laser field, as for atoms. The phase transformations can be induced
by a Gaussian laser beam with the intensity profile
\begin{equation}
I(r, z) = \frac{I_0}{1+
\frac{z^2}{z_R^2}}\exp\left[-\frac{2r^2}{w_0^2\left (1+
\frac{z^2}{z_R^2}\right) }\right] \ ,
\label{gaussian-intensity-profile}
\end{equation}
where $I_0$ is the light intensity at the beam center, $r$ is the
radial distance from the center axis of the beam, $z$ is the axial
distance from the beam center,  $z_R=\pi w_0/\lambda$ is
Rayleigh range, $w_0$ is the beam waist and $\lambda$ is the wavelength.
With the 1D molecular array arranged along the $z$-axis, the laser field intensity
can be made to vary linearly along the array,
\begin{equation}
I(r=na, z=0; t) \approx [I_{c} + n I_1 ] \, \sin^{2}(\pi t/T) \; \;\; (0 < t <
T) \ , \label{intensity-varying}
\end{equation}
where $I_{c}$ is the intensity at the center of the wave packet.
This can be achieved if $z_0 = z_R/\sqrt{3}$ and ${\sigma}_x^{(2d)}
a\lesssim 0.5 z_R$, where $z_0$ is the distance between
the center of the wave packet and the beam center, and ${\sigma}_x^{(2d)}$ is the width (in the coordinate representation) of the two-dimensional wave packet. Using Eqs. (\ref{phase}), (\ref{DressedEnery-Sum}) and (\ref{gaussian-intensity-profile}), we estimate the momentum kick by such a pulse as $
\delta = -\sqrt{3}T I_{0} (\alpha_{\|}-\alpha_\perp) / 80 z_R
$.
The results presented in Figure 1 were obtained for a 1D array of LiCs molecules on an optical lattice with $a=400$ nm and the external perturbation
given by the laser field pulse (\ref{intensity-varying}) with parameters $I_c$ and $I_1$ derived from Eq. (\ref{gaussian-intensity-profile}) with $z_0 = 45\; \mu$m and
$z_R=73.8\; \mu$m. The numerical results deviate from the analytical prediction for $\delta$ by less than 7 \%.

The Gaussian intensity profile (\ref{gaussian-intensity-profile}) can be used also to implement the quadratic phase transformations needed for focusing of collective excitations. 
To achieve this, a 2D molecular array must be arranged in the $z=0$ plane,
with the  $x$-axis defined to be along the polarization direction of a linearly polarized field.
If the dimension of the molecular array is smaller than one third of the
beam waist, the Gaussian intensity profile in
Eq. (\ref{gaussian-intensity-profile}) can be approximated
by
\begin{equation}
I(r=na, z=0; t) \approx I_{0}\left[1- \frac{2(n_x^2 +
n_y^2)a^2}{w_0^2}\right]. 
\label{quadratic-profile}
\end{equation}
This is a concave quadratic intensity
profile which can be used to focus a wave packet  in a system with negative couplings $\alpha$ (see Section 3).

\section{Control of energy transfer in dipolar systems}

Dipolar interactions play a central role in the study of long-range interaction effects using ultracold systems \cite{our-njp-review}.
While, in general, the coupling constant $\alpha$ in Eq. (\ref{ham}) can be determined by a variety of interactions, the dominant contribution to $\alpha$ for atoms and molecules on an optical lattice is determined by the matrix elements of the dipole - dipole interaction. It is therefore particularly relevant to discuss the specifics of energy transfer in systems with dipolar interactions. 

The dipolar interactions are long-range and anisotropic. The long-range character of the dipolar interactions 
manifests itself in the modification of the exciton dispersion (\ref{Eexc}). While Eq. (\ref{Eexc}) is valid for a system 
with nearest neighbour couplings only, higher-order couplings in the case of $\alpha(n-m) \propto 1/(n-m)^3$ 
modify the exciton dispersion leading to a cosine-like, 
but non-analytic dispersion relation, both in 1D  and 2D. 
To investigate the effect of this nonanalyticity in dispersion curve, we have performed a series of calculations with the long-range couplings neglected 
after a certain lattice site separation $n-m$ for the 1D system. The results become  converged (to within 0.2 \%) when
 each molecules is directly coupled with 20 nearest molecules. While the calculations with only the nearest neighbor
 couplings are in good agreement  with the analytical predictions given by Eqs. (\ref{focus-time}) and (\ref{x-focusing-Gauss}), the full calculations reveal that long-range 
couplings somewhat decrease the focusing efficiency. The long-range couplings also decrease the focusing time, 
by up to a factor of 2. The dynamics of collective excitations leads to interference oscillation patterns clearly 
visible in panels b and d of Figure \ref{focusing-1d}. These oscillations are much less pronounced when all but nearest
 neighbor couplings are omitted. The numerical results of Figures 1 - 3 are particularly important
 because they demonstrate that the phase transformations introduced in the present work are effective  for 
systems with dipolar interactions.

The anisotropy of the dipolar interactions can be exploited for controlling energy transfer in dipolar systems by varying the {\it orientation} of a dressing external DC electric field. 
For example,  for polar molecules on an optical lattice,  
the matrix elements  $\alpha(n-m) = \langle e_{n} , g_{m} | {V}_{\rm dd}(n-m) | g_{n}, e_{m} \rangle$ depend not only on the choice of
 the states $|g\rangle$ and $|e\rangle$, but also on the magnitude and orientation of an external dc electric field
\cite{biexcitons, felipe}. Since the value of $\alpha$ determines the exciton dispersion (\ref{Eexc}),
 the exciton properties can be controlled by varying the angle $\theta$ between the intermolecular axis
and the applied DC field. This is illustrated in Figure 4. 

The calculations presented in Figure \ref{control-exciton} are for a 1D array of LiCs molecules in a lattice with  $a=400$ nm.
As before, $|g\rangle$ is the absolute ground state of the molecule and $|e\rangle$ is the rotationally excited state
 that adiabatically correlates
with the rotational state $|J=1, M_J = 0 \rangle$ in the limit of vanishing electric field. 
The upper panel of Figure 4
shows that the angle $\theta$ between the electric field vector and the molecular array axis
determines the sign and magnitude of $\alpha$, and therefore the shape of the dispersion
curve. This enables
control over the sign and magnitude of the group velocity of an
excitonic wave packet containing contributions with $k\neq 0$. Dynamically tuning $\theta$,
one can propagate a localized excitation to different parts of the
lattice, as shown in Figure \ref{control-exciton}b.

In a 2D lattice, the intermolecular interactions depend on an additional azimuthal angle $\phi$ that describes
 the rotation of the electric field axis around the axis perpendicular to the lattice. The numerical calculations 
presented in Figure \ref{control-2d-wavepacket} show that the energy flow in two dimensions can be  
controlled by varying both $\theta$ and $\phi$. 
In addition to the phase transformation discussed earlier, this allows for a dynamical energy transfer in quantum 
many-body systems with anisotropic interparticle interactions.

\begin{figure}[ht]
\centering
\includegraphics[width=\linewidth]{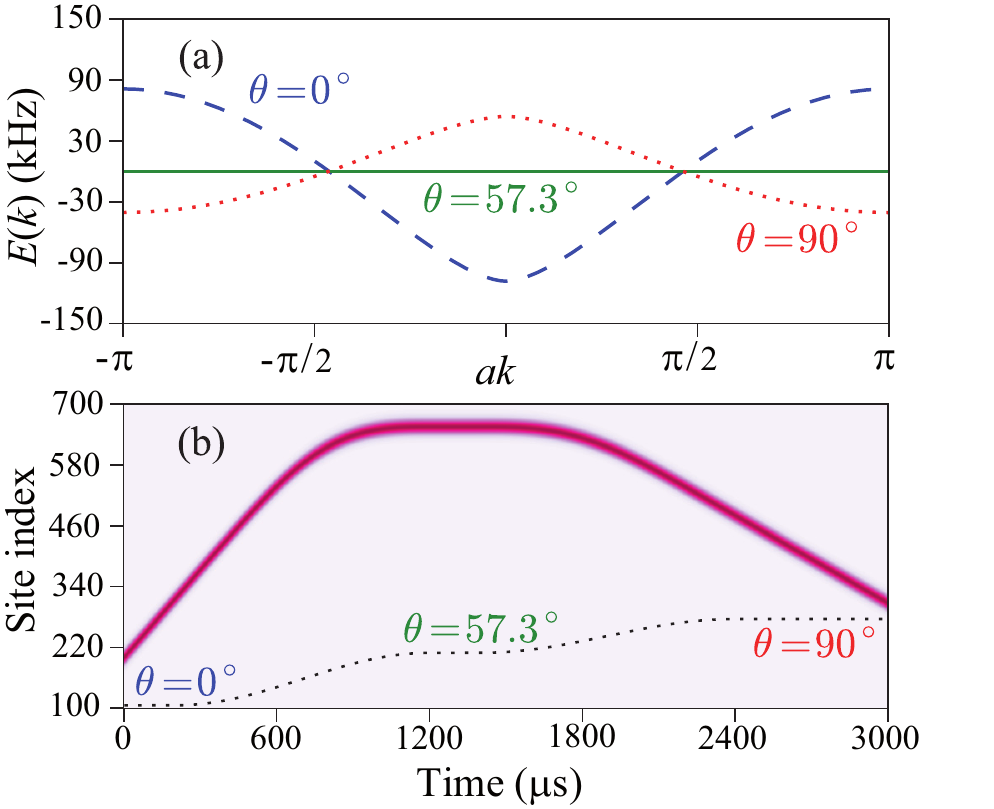}
\caption{ (Color online) (a) Exciton dispersion curves for a 1D
ensemble  of diatomic molecules on an optical lattice for
different angles $\theta$ between the direction of the external DC
electric field and the axis of the molecular array.  In 1D, the coupling $\alpha\propto (1/3 - \cos^2\theta)$.  (b)
Propagation of a wave packet centered at $ak =-\pi/3$ controlled
by tuning the electric field direction. Thin dashed line depicts
the corresponding angle variations with time. }
\label{control-exciton}
\end{figure}

\begin{figure}[h]
\centering
\includegraphics[width=\linewidth]{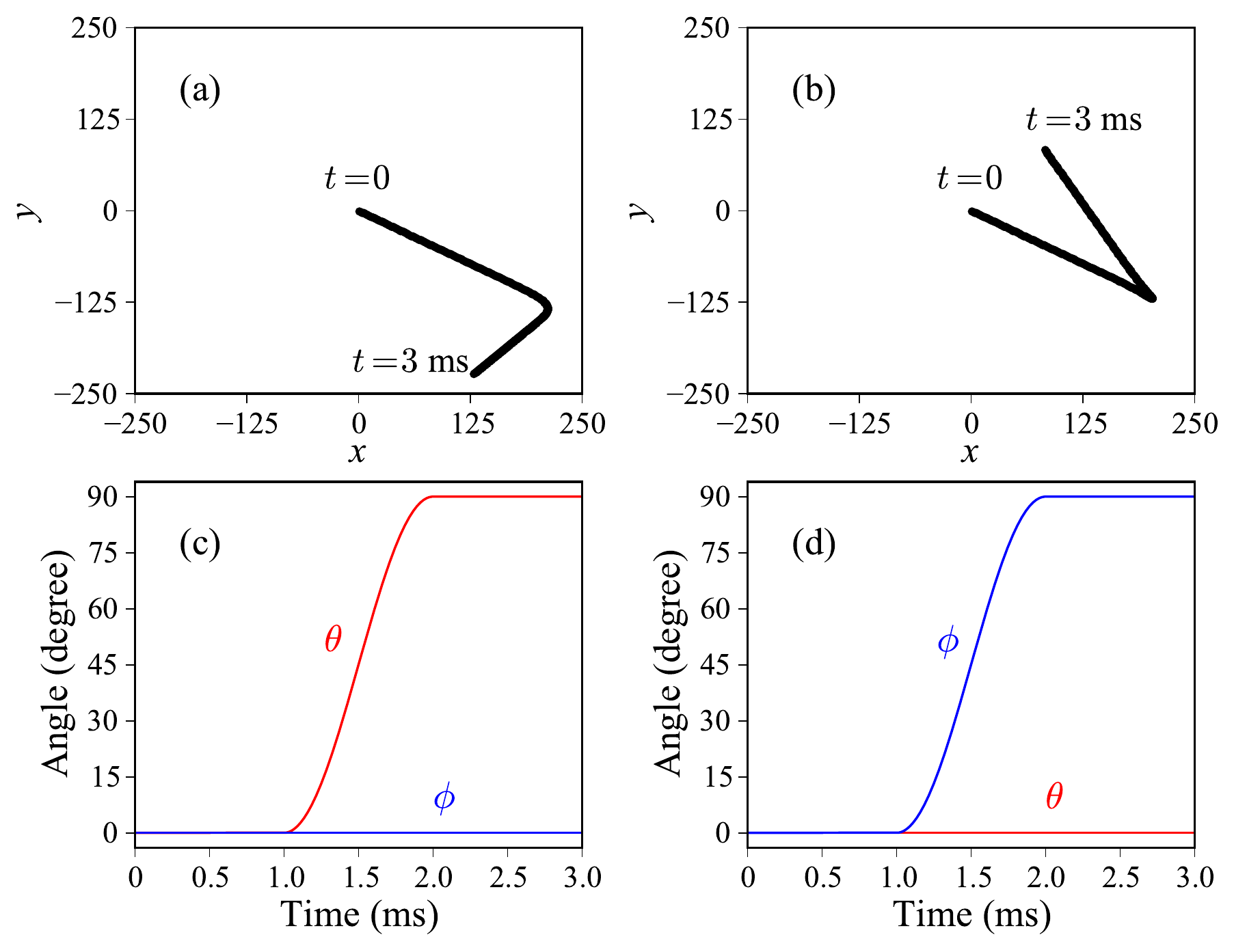}
\caption{ (Color online) (a) and (b) show the trajectories of the center of an exciton wave packet in a 2D lattice during
 the time from 0 to 3 ms; (c) and (d) represent the changing of the dressing DC field orientation $(\theta, \phi)$ associated with (a) and
(b) respectively. The initial wavepacket is a 2D Gaussian distribution centered around $ak_x=ak_y=\pi/2$ and has a width of $\sim$60 lattice sites in coordinate space. The magnitude of the DC field is fixed to
6 kV/cm while its direction is changing. The calculations are done for a 2D array of  LiCs molecules in a lattice with
$a=400$ nm.
 }
\label{control-2d-wavepacket}
\end{figure}

\section{Energy transfer in the presence of vacancies}

While experiments with ultracold atoms have produced a Mott insulator phase with  99\%  of lattice sites filled \cite{atom-mott1, atom-mott2, atom-mott3},
the latest experiments with molecules yield lattice-site populations about  10\% \cite{Ye-arrays-PRL12}.
Multiple experiments are currently underway to produce a Mott insulator phase of polar molecules with close to the full population of the lattice.
However, lattice vacancies may be unavoidable in the best experiments. In this section, we examine the effect of vacancies on the possibility of focusing collective excitations
to a desired region of the lattice by the phase transformations discussed in Section 3. For concreteness, we perform calculations for the system described in Section 4,
namely a 2D array of LiCs molecules on a square optical lattice with $a=400$ nm.


To explore the effect of vacancy-induced interactions, we performed simulations for
different vacancy numbers using the same parameters for molecule-field and inter-molecular interactions as
in the calculations presented in Figure \ref{focusing-2d}b. For each vacancy concentration, we carried out 48
calculations with random distributions of empty lattice sites. The quadratic phase transformations are applied,
as described in Section 3, in order to focus the collective excitation at time $t_{\ast}$ to
the molecule in the middle of the 2D array.

Vacancies disturb the translational symmetry of the system and produce an effective disordered potential that tends to
localize collective excitations \cite{perez-rios2010}.
 Because the natural time evolution of the wave packet in a disordered potential may lead to enhancement of the probability in certain
regions of the lattice, it is necessary to distinguish the effect of the vacancy-induced localization and the effect of the focusing phase
transformation. To quantify these two effects, we define two factors:
the enhancement of the probability at the target molecule with respect to the initial value,
\begin{equation}
\eta = \frac{p' (t=t_{\ast})}{p(t=0)},
\label{eta}
\end{equation}
and the ratio of the probability to find the excitation on the target molecule with ($p'$) and without ($p$) the focusing
phase transformation,
\begin{equation}
\chi = \frac{p' (t=t_{{\ast}})}{p (t=t_{ {\ast}})}.
\end{equation}
The time $t_*$ is the focusing time predicted in Section 3 for an ideal, vacancy-free system.
 The quantity $\eta$ illustrates the actual enhancement of the probability to focus a collective excitation,
while the quantity $\chi$ illustrates the effect of the focusing phase transformation.
Figure 6 presents the values of $\eta$ and $\chi$ as functions of the vacancy concentration.
It illustrates two important observations. First, the disorder potential with vacancy concentrations $> 20$ \%
renders the phase transformation uneffective. In the presence of strong disorder, the dynamics of the system is entirely
determined by the disorder potential and the energy transfer becomes highly inefficient (however, see Section 7). On the other hand, vacancy
concentrations of less than 10 \% appear to have little effect on the efficacy of the focusing phase transformation.

Our calculations indicate that the focusing time may be somewhat modified by the disorder potential, even if the
 concentration of vacancies is less than 10 \%.  Figure 7 depicts the excitation wave functions
 at the time of the maximal enhancement on the target molecule, chosen as molecule (71,71).  Figure \ref{focusing-with-vacancy} shows that
despite the presence of multiple vacancies, the focusing transformation enhances the probability to find the
excitation on the target molecule by 16 times.


\begin{figure}[ht]
\centering
\includegraphics[width=\linewidth]{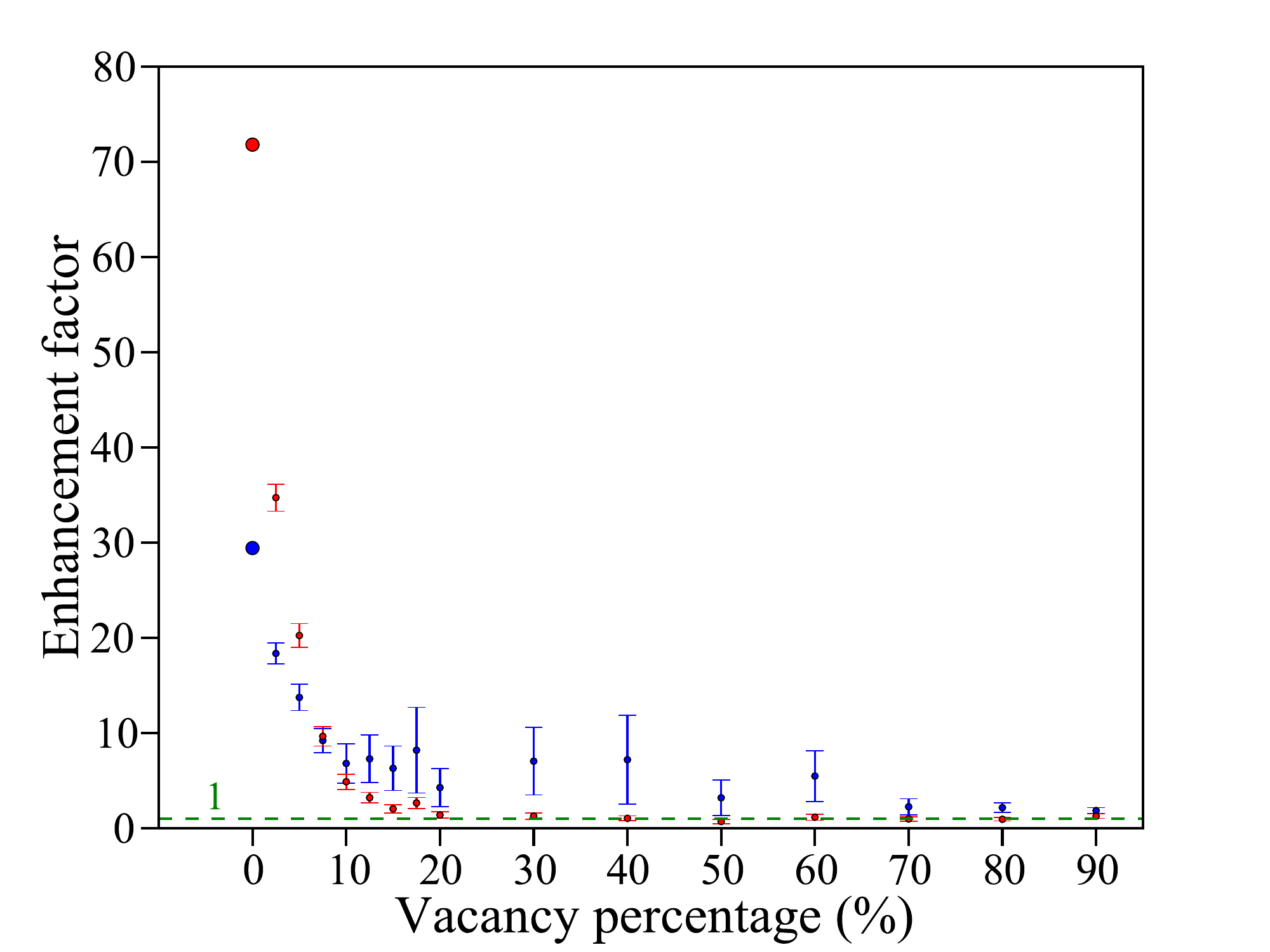}
\caption{ (Color online) Enhancement factors $\eta$ (red symbols) and
$\chi$ (blue symbols) as functions of vacancy percentage in a 2D lattices. See text
for the definitions of $\eta$ and $\chi$. The error bars are for 95\% of confidence interval.} 
\label{enhancement-vs-vacancy}
\end{figure}


\begin{figure}[h]
\centering
\includegraphics[width=\linewidth]{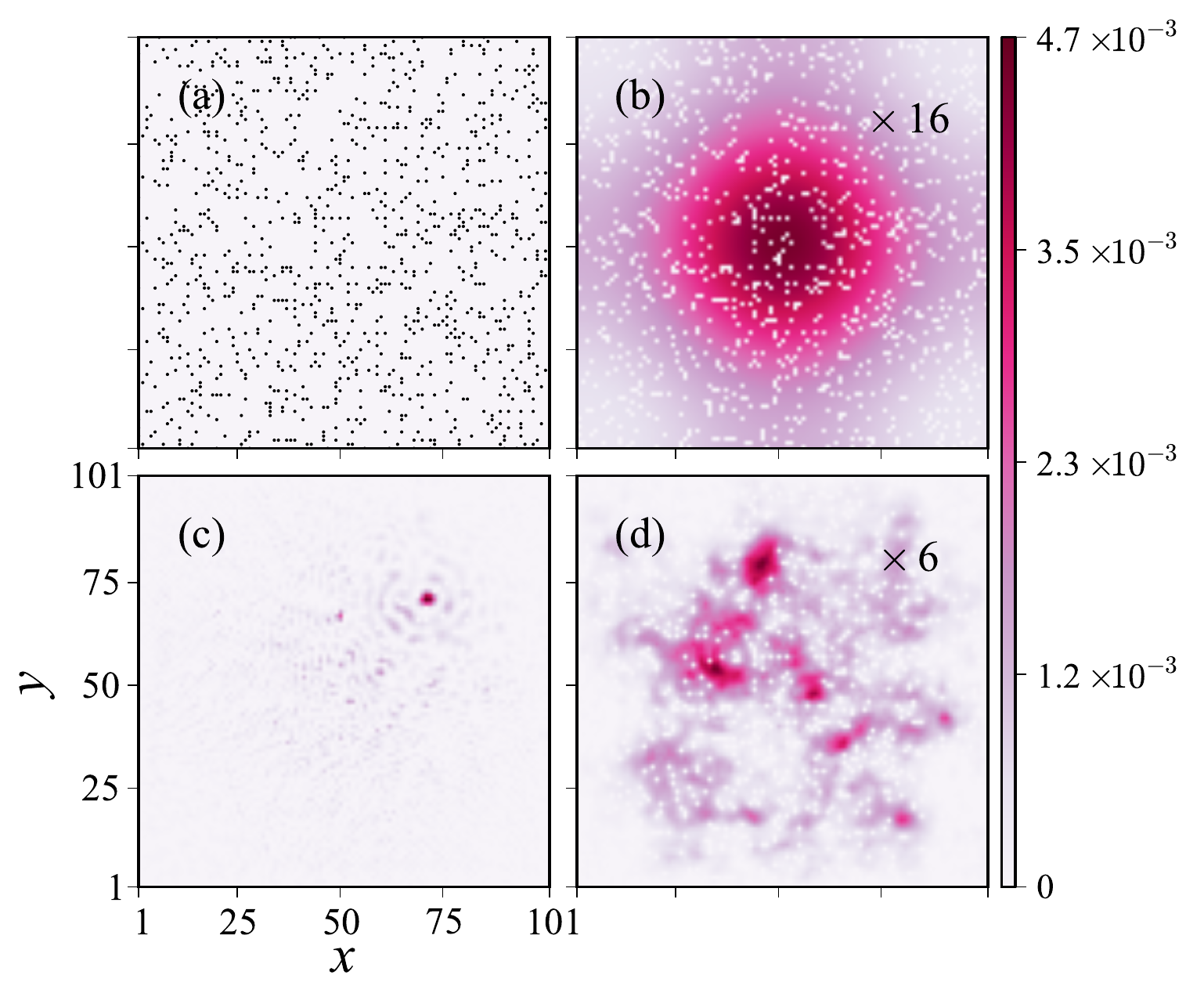}
\caption{ (Color online) Time snapshots of a collective excitation in a 2D
array with a vacancy concentration of 10 \%  (a) The distribution of the vacant sites; (b) The initial probability distribution of the excited state; (c) The probability distribution of the excitation at the
focusing time when the focusing scheme is applied. The focusing
time is defined to be the time when the probability at the target
molecule (71, 71) reaches maximum. (d) The probability
distribution of the wave function at the focusing time when the
focusing scheme is not applied. The calculations are performed
with the same parameters as in Figure \ref{focusing-2d}. The
probabilities in (b) and (d) are enhanced by 16 and 6
respectively. } \label{focusing-with-vacancy}
\end{figure}

\section{Focusing in the presence of strong disorder}


Although the focusing method demonstrated in Sections 3 and 6 appears to be robust in the presence of a disorder potential induced by a small concentration of vacancies, it is important for practical applications to also consider controlled energy transfer in quantum arrays under a strong disorder potential. To consider focusing in a strongly disordered system, 
we employ an analogy with the {``transfer
matrix'' methods for focusing of a collimated light beam in opaque
medium \cite{opaque-1, Gigan-TMeasure-PRL10, Mosk-NPhot10, Cizmar-NPhot10, Silberberg-11, Chatel-Focusing-11, Lagendijk-Focusing-11, zhenia-11, cui-11, kim-11}}.

In optics, a collimated laser beam passing through an opaque medium results in a random
pattern of speckles arising from random scattering of light inside the medium
\cite{RandomWave-books}. Likewise, the random distribution of empty sites in an optical lattice with molecules 
scatters the exciton wavepackets, resulting in a completely random excited state. 
However, in optics, the randomness of the scattering centers inside the opaque medium can be compensated for by
shaping the incident wavefront with a spatial light modulator such
that the contributions from various parts of the medium can add
constructively upon exit from the medium, producing a focus. 
We suggest that the same can be achieved with a many-body system on a lattice by 
separating the entire lattice into multiple blocks and applying proper phase transformations
to those individual blocks.

%

The initial state for an ensemble of molecules on a lattice with multiple vacancies can be written as 
\begin{equation}
|\psi(t=0)\rangle = \sum_{i}c_i(t=0)|i\rangle \ , \label{planewave-initial}
\end{equation}
where 
\begin{eqnarray}
| i \rangle = |e_i\rangle \prod_{j\neq i} |g_j\rangle
\end{eqnarray}
and the indexes $i$ and $j$ run over all occupied sites. After a long evolution time $T$, the probability amplitude for the excitation to reside on a particular target molecule is given by  
\begin{equation}
c_{o}(T) = \sum_{i} U_{o, i}(T)c_i(t=0) \equiv \sum_i c_{oi}(T), 
\label{site-contribution}
\end{equation}
where $U_{o,i}(t) = \langle o| \exp[-iH_{\rm exc}t] | i \rangle$ is a matrix element of the time evolution operator.
{In a disordered system, the transfer coefficients $U_{o, i}$
are not a-priori known and depend on the
disorder potential. The {phasors} $c_{oi}(T)$ have quasi-random
amplitudes and phases. While the amplitude of each phasor cannot be controlled experimentally, 
their phases are controllable via the phases of the coefficients $c_i$ at $t=0$, which can be tuned using the phase-kicking
transformations introduced above.} To achieve the highest probability
at the target molecule, it is necessary to ensure that the contribution
$c_{oi}=U_{o, i}(T)c_i(t=0)$ from every site $i$ has the same
phase so that they add up constructively.

In a practical implementation, it may be difficult to control the phase of each molecule in each
individual site. It may be more desirable to work with blocks of several lattice sites. Assuming that the entire array of molecules 
can be divided into $M$ blocks, each containing many molecules, and that the blocks can be perturbed individually, the excitation probability amplitude at the target molecule at time $T$ is 
%
\begin{equation} c_{o}(T) =  \sum_{\gamma=1}^M c_{\gamma}(T)
\label{blocks-contribution}
\end{equation}
where
\begin{equation} c_\gamma(T) \equiv |c_\gamma| e^{i\phi_\gamma}= \sum_{i\in
 \gamma} U_{o, i}(T)c_i(t=0) \ .
\label{inblock-contribution}
\end{equation}
%
%
This equation implies that the contributions
from different blocks can be made to interfere constructively by adding
 a phase $\exp(-i \phi_{\gamma})$ to each occupied site in block
 $\gamma$. For $M$ blocks in the array and quasi-random evolution matrix, simply setting all the phases equal
 must lead to $\sim M$-fold increase of the excitation probability at the target molecule, as compared to
 a sum of $M$ quasi-random phasors in Eq. (\ref{blocks-contribution}) \cite{opaque-1}.

Similarly to optics, the phases $-\phi_\gamma$ which must be added in each block, can be found experimentally provided
that the same (or similar) realization of disorder persists in a series of
trials. A straightforward optimization would scan through the
strengths of  phase kicks applied to different blocks. In each
experiment one would measure the excitation probability at the
target molecule $|c_o(T)|^2$, e.g. via resonance fluorescence from
the target molecule at the end of the experiment. More
sophisticated optimization techniques, aimed at fast focusing
multi-frequency light in optical systems, are currently under
rapid development \cite{Gigan-TMeasure-PRL10, Mosk-NPhot10, Cizmar-NPhot10, Silberberg-11, Chatel-Focusing-11, Lagendijk-Focusing-11, zhenia-11, cui-11, kim-11}.


For a proof-of-principle calculation, we consider a 2D lattice of size 101$\times$101 with 60\% of sites vacant and each non-vacant site occupied by a single LiCs molecule. 
Due to time reversibility of the time
evolution operator $U(T)$, 
\begin{equation}
|c_\gamma | \exp(-i \phi_{\gamma}) = \left[ \sum_{j=1}^{n} U_{o, j}^\gamma (T) c_j^{\gamma}(0)\right]^* 
=\left[ \sum_{j=1}^{n} U_{j, o}^\gamma (-T) c_j^{\gamma}(0)\right]^* \ .
\end{equation}
The matrix element $U_{j, o}^\gamma (-T)$ can be
calculated by performing a backward time propagation starting from a
local excitation at site ``$o$'' and calculating the coefficient
$c_j(t)$ at time $-T$. Alternatively, one can propagate the evolution equations forward in time, finding $c_j(T)$: 
Since the Hamiltonian (1) is real, its eigenfunctions are real, and the evolution matrix $U$ is symmetric, $U_{o, j} = U_{j, o}$. Thus we find
\begin{equation}
c_j(T) = \sum_{i}U_{j, i}^\gamma (T) c_i(0) = U_{j, o}^\gamma (T) \ ,
\end{equation}
since $c_{o}(0) = 1$ and all other coefficients are zero. For a completely delocalized initial state, we assume that all coefficients
in Eq. (\ref{planewave-initial}) are equal, so that  the phases $\phi_\gamma$ required for
block $\gamma$ are
\begin{equation}
 |c_\gamma|\exp(- i \phi_{\gamma})  = \left[ \sum_{j}  c_j(T) \right]^* \ ,
\end{equation}
where the index $j$ runs over all  occupied sites in block $\gamma$.
Figure \ref{t-matrix-focusing} shows that this choice of phases leads to effective focusing of the collective excitation in a strongly disordered system. 

\begin{figure}[ht]
\centering
\includegraphics[width=\linewidth]{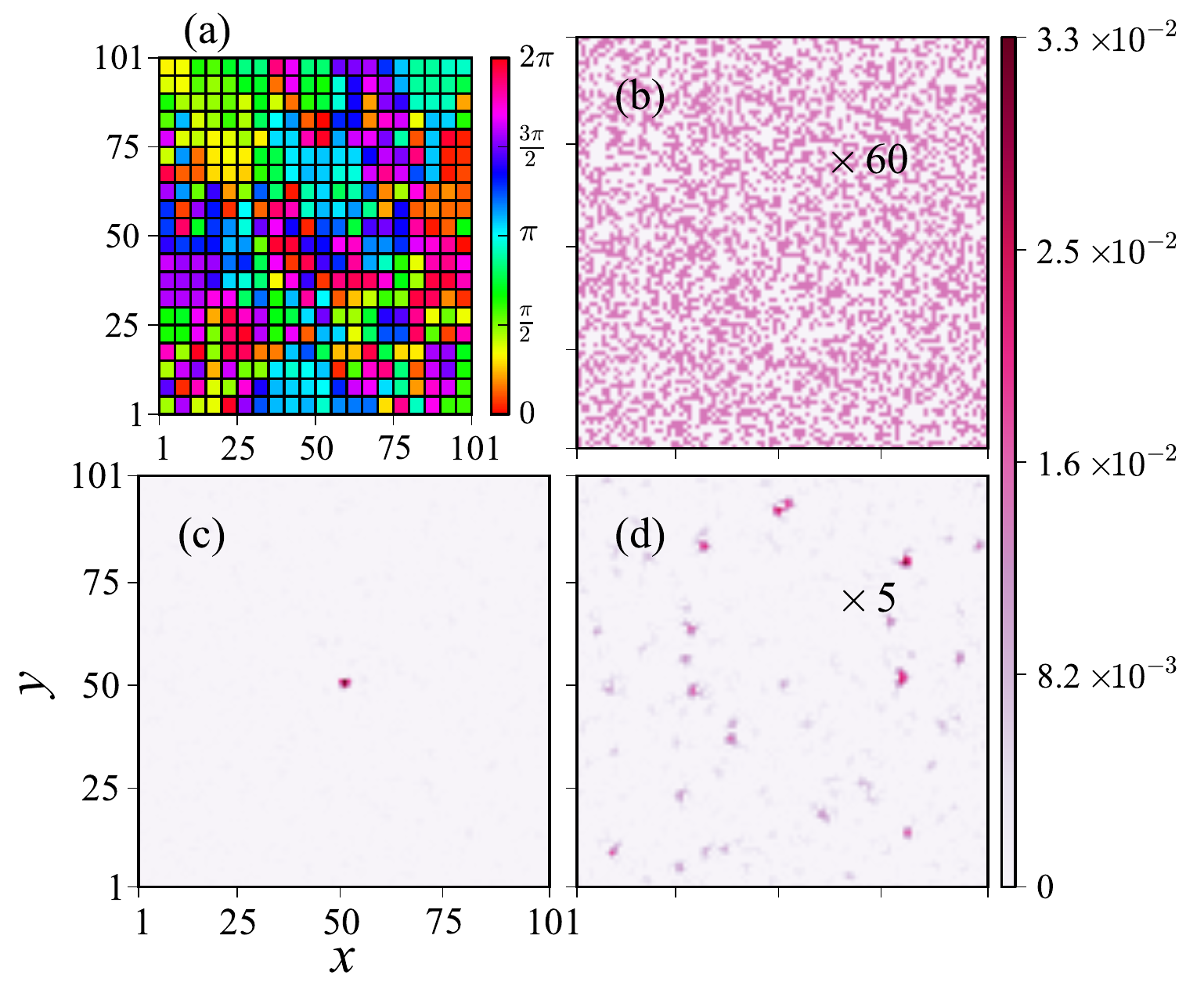}
\caption{ (Color online) Focusing of a collective excitation in a strongly disordered system with 60\% of lattice sites unoccupied.  Panel (a) shows different phases applied to
different blocks of the lattice before the time evolution.
(b) The initial probability distribution of the excited state. (c) The probability distribution of the excited state at the
focusing time $T = 3$ ms with the phase transformation depicted in panel (a) before the time
evolution. (d) The probability distribution of the excited state at the focusing time $T = 3$ ms with no phase transformation
applied. The calculations are performed
with the same parameters as in Figure \ref{focusing-2d}. The
probabilities in (b) and (d) are enhanced by 60 and 5,
respectively. } \label{t-matrix-focusing}
\end{figure}

To illustrate the efficiency of the focussing method described above, we have carried out a series of calculations 
with different vacancy concentrations. For each vacancy concentration, we performed 48 calculations with random
 distributions of empty lattice sites. The phase transformations are calculated individually for each random 
distribution of vacancy sites as described above. The results are shown in Figure \ref{enhancement-vs-vacancy-t-matrix}. 
As can be seen, the transformations proposed above are effective for vacancy concentration $<$ 70\%. At higher 
concentrations of vacancies, the excited states become strongly localized and immobile. The focusing efficiency at
vacancy concentrations 10\% and 20\% appears to be higher than that in the absence of vacancies, which we attribute to the effect of 
the boundaries.
\begin{figure}[ht]
\centering
\includegraphics[width=\linewidth]{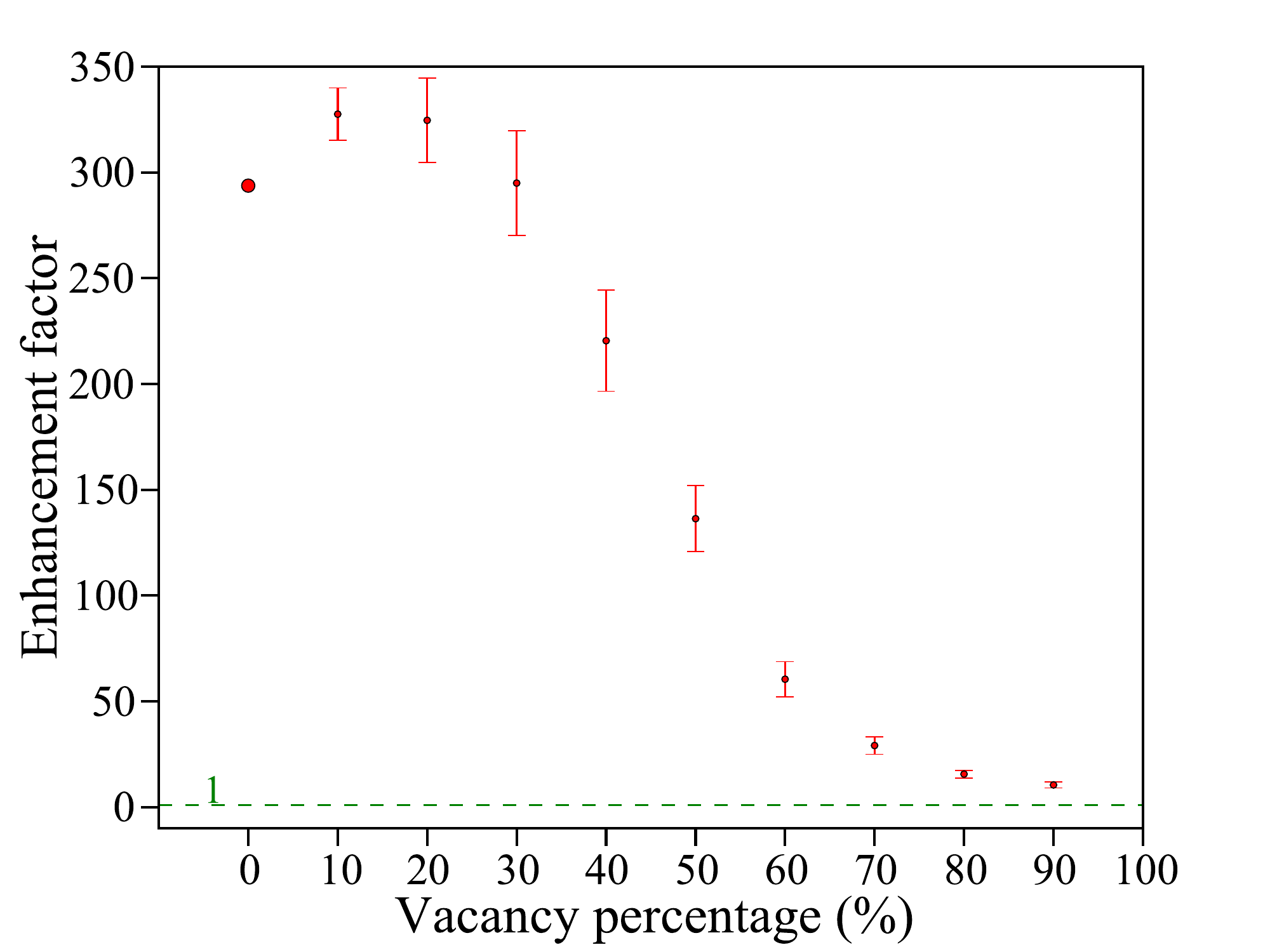}
\caption{ (Color online) Enhancement factors $\eta$ (red symbols) as a function of vacancy percentage in a 2D lattices. 
$\eta$ has the same definition as in  Eq. (\ref{eta}) except the time $t_*$ is arbitarily chosen to be 4 ms here. 
 The error bars are for 95\% of confidence interval.} 
\label{enhancement-vs-vacancy-t-matrix}
\end{figure}

\section{Conclusion}

We have proposed a general method for controlling the time evolution of quantum energy transfer in ordered 1D and 2D arrays of coupled monomers.
Any elementary excitation in an aggregate of coupled
monomers can be represented as a coherent superposition of Frenkel
exciton states. We propose shaping the exciton wave
packets using nonadiabatic perturbations that temporarily
modulate the energy levels of the monomers leading to monomer-dependent linear phase transformation and a
displacement of the wave packets in the wave vector representation.
This, combined with the possibility of focusing a collective excitation on a particular part of the lattice by a quadratic phase transformation and with the directed propagation of collective excitations,
allows for control of energy transfer in the lattice.

We have presented numerical calculations for an ensemble of polar molecules trapped on an optical lattice that demonstrate the feasibility of
both momentum-shifting and focusing of collective excitations by applying external laser fields, with parameters that can be easily achieved in the laboratory.
We also investigated the effect of disorder potential arising from incomplete population of the lattice. Our results show that the phase transformations
 leading to focusing of collective excitations on different regions of a 2D lattice remain effective in the presence of vacancies
with concentrations not exceeding 10 \%. For systems with larger concentrations of vacancies and affected by strong disorder potentials, we propose an alternative procedure based on engineering constructive interference of the wave function contributions arising from difference parts of the lattice.

The momentum-shifting technique proposed here can be used to protect collective excitations of ultracold atoms from spontaneous emission. 
 The spontaneous decay processes, which in the case of an ordered many-body
system must satisfy both the energy and wave vector conservation rules, can be restricted by shifting the exciton
 wave packets to a region of the dispersion curve, where the wave vector conservation cannot be satisfied.
If performed faster than the spontaneous emission time, such phase transformations should create collective
excitations with much longer lifetimes, which opens a variety of new applications for ultracold atoms on an optical
lattice.

Control over excitation transfer is needed for creating networks
of quantum processors where information is transmitted over large
distances with photons and stored in arrays of quantum monomers
via one of the quantum memory protocols \cite{Lvovsky-Qmemories}.
Momentum kicking can be used for information transport within a
single array. Focusing excitonic wave packets enables local
storage of information, while directed propagation combined with controlled interactions of multiple
excitons \cite{biexcitons} or excitons with lattice impurities
\cite{zoller-atom-transistor} may be used to implement logic
gates. Controlled energy transfer in molecular arrays may also be used
for the study of controlled chemical interactions for a class of
reactions stimulated by energy excitation of the reactants.
Directing quantum energy to a particular lattice site containing
two or more reagents can be used to induce a chemical interaction
\cite{pccp}, an inelastic collision or predissociation
\cite{wallis-krems}  with the complete temporal and spatial
control over the reaction process.

Finally, the present work may prove to be important for simulations of open quantum systems. 
We have recently shown \cite{felipe, felipe-arxive-polaron} that
the rotational excitations of ultracold molecules in an optical lattice can,
by a suitable choice of the trapping laser fields, be effectively coupled to lattice phonons.
The exciton - phonon couplings can be tuned from zero to the regime of strong interactions
\cite{felipe, felipe-arxive-polaron}. The possibility of shaping (accelerating, decelerating and focusing)
collective excitations as described in the present work combined with the possibility of coupling these excitations
to the phonon bath opens an exciting prospect of detailed study of controlled energy transfer in the presence of a
controllable environment. Of particular interest would be to study the effect of the transition from a weakly coupled
Markovian bath to a strongly coupled non-Markovian environment on energy transfer with specific initial parameters.

We note that  the effect of site-dependent phase transformations on quantum transport was independently considered 
in Ref.  \cite{zimboras2012} from the point of view of time-reversal symmetry breaking. The authors of Ref.  \cite{zimboras2012} propose 
an experimental realization based on ions in a linear Paul trap. Their method relies on the possibility of tuning time-dependent phases, 
leading to new effects. The present work and Ref. \cite{zimboras2012} should be considered complementary.

\section*{Acknowledgment}

This work was supported by NSERC of Canada
and the Peter Wall Institute for Advanced Studies.

\section*{References}

\bibliographystyle{unsrt}

\end{document}